\documentclass[prl,showpacs,preprint]{revtex4-1} 
\usepackage{xspace} 
\usepackage{makecell} 
\usepackage{siunitx, mhchem}
\usepackage{color,soul}
\usepackage{graphicx}
\usepackage{bm}        
\usepackage{amssymb}   
\usepackage{amsmath,mathtools} 
\usepackage{breqn} 
\usepackage{multirow}
\usepackage{array}
\usepackage{booktabs}
\usepackage[export]{adjustbox}
\usepackage{textcomp}
\usepackage[normalem]{ulem} 
\usepackage{gensymb}
\usepackage[autostyle]{csquotes}
\newcommand{\mos}{MoS$_2$\xspace}

\newcommand{\sio}{SiO$_2$/Si\xspace}
\begin{document}

	
	
	\title{Effect of strain and sulfur vacancies on the luminescence and valley polarization properties of CVD grown monolayer \mos films}

	\author{Poulab\ Chakrabarti}
	\email{poulab007@gmail.com}
	\author{Faiha\ Mujeeb}
	\author{Subhabrata\ Dhar}
	\affiliation{Department of Physics$,$ Indian Institute of Technology Bombay$,$ Mumbai 400076$,$ India}

	\begin{abstract}
		 
Using temperature dependent photoluminescence (PL), polarization resolved PL and Raman spectroscopy, we investigate the effect of \textit{in situ} vacuum annealing as well as the relaxation of strain on the luminescence and the valley polarization properties of large area strictly monolayer (1L)-\mos, grown on sapphire and \sio substrates by a microcavity based chemical vapor deposition (CVD) technique. The study shows that the strain as well as the physisorption of air molecules at the sulfur vacancy ($V_S$) sites play key roles in governing the optical quality of CVD grown 1L-\mos. Removal of air molecules from the $V_S$ sites  enhances the relative strength of the A-exciton/trion transition as compared to the broad luminescence (BL) band arising from those defects at low temperatures. It has also been found that such removal  helps in improving the valley polarization property of the film. Relaxation of biaxial tensile strain, which has been achieved by post growth transferring of 1L-\mos film from the sapphire to a \sio substrate by a polystyrene assisted transfer process, is also found to be helpful to get back the high polarization character ($\sim$80\%) of the valleys. The study further shows that the transfer process not only facilitates the removal of physisorbed air molecules from the $V_S$ sites but also puts in place a long lasting capping layer on \mos that shields the film from reacting with air and hence enhances the relative yield of A-exciton/trion transition by suppressing the BL transition. The study thus creates an opportunity to use CVD grown large area 1L-\mos for the development of  optoelectronic as well as valleytronic  devices for practical applications for the future.
		 
	\end{abstract}
	
	
	\maketitle
	

Single layer (1L) Molybdenum disulfide (\mos) has generated significant interest in the last decade due to its novel electronic and optical properties. The material offers several fascinating features, such as direct band gap at the Brillouin zone K-points\cite{mak1,bs}, large exciton binding energy due to reduced dimensionality\cite{mouri_Ch}, existence of many body bound states, e.g., trions\cite{mak_trion} as well as biexcitons\cite{mai_manybody} and most importantly spin-valley coupling resulting in valley polarization (VP)\cite{mak2,cao_vp,zeng_vp1} that can be exploited to inject spin polarized carriers and excitons for future spintronic and valleytronic devices \cite{zeng_vp1,xiao_vp2,Kawakami1}. Mechanical exfoliation (ME) and chemical vapor deposition (CVD) are the two most popular techniques to obtain 1L-\mos. However, the two techniques offer contrasting benefits and pitfalls. While, ME can ensure the premium crystal quality of the film by choosing high-grade bulk parent material, getting large area monolayer coverage is impossible in this technique\cite{budania_me1,magda_me2}. Whereas, CVD route can synthesize continuous 1L-\mos extending up to a few cm$^2$ area\cite{baek_CVD,pkm} providing the necessary platform for the fabrication of large scale integration of devices. Unfortunately however, a large density of sulfur vacancy ($V_S$) related defects\cite{tailoring,surrente_CVD_vp}, grain boundaries\cite{najmaei_gb} and residual  strain\cite{amani_dubey1} are invariably present in these films that degrade the device performance. These $V_S$ related defects, which are known to act as shallow donors in \mos\cite{vinod_vs,therm}, are shown to be passivated due to physisorption of  ambient oxygen-molecules at the vacancy sites leading to the significant reduction of conductivity of 1L-\mos\cite{therm,amani_dubey1,tailoring}. Photoluminescence (PL) studies on 1L-\mos films often show the presence of a broad luminescence (BL) peak, which is also attributed to the adsorption of ambient oxygen and nitrogen at the $V_S$-sites\cite{tongay_defect}. In fact, there are several experimental studies on the role of surface adsorbates, substrates, strain on the luminescence characteristics of both ME\cite{tongay_defect,tongay_defect1,nan_defect,chow_defect,zhu_vp_strain,conley_raman,frisenda_biaxial} and CVD\cite{amani_dubey1,dubey_strain2,gogoi_ad3,sun_defect,rahul_ad2} grown 1L-\mos. However, very little has been done to experimentally explore the effect of these factors on the valley polarization property of 1L-\mos. These studies are crucial for the development of future valleytronic devices on 1L-\mos platform. Such studies become even more important in case of CVD grown 1L-\mos films as these are always grown on a substrate, which can introduce biaxial strain in the layer and in addition, CVD grown layers posses much higher density of $V_S$ defects as compared to exfoliated films\cite{surrente_CVD_vp}. Zhu \textit{et al.} have studied the effect of uniaxial tensile strain on the VP properties at room temperature for 1L-\mos obtained by ME technique and found a reduction of VP with the increase of the strain\cite{zhu_vp_strain}. Saigal \textit{et al.} have reported high value of polarization anisotropy of the A-excitonic PL feature for CVD grown 1L-\mos on \sio substrates, after removing surface residues through high temperature annealing under ultra high vacuum condition\cite{raman2}. But a systematic effort is needed to understand the underlying mechanism. Moreover, a stable and sustainable method has to be developed to avoid the degradation of the property due to exposure to the ambiance. Surrente \textit{et al.} have reported charge transfer mediated VP of 1L-MoSe$_2$ flakes sandwiched between two more 1L-\mos flakes\cite{surrente_CVD_vp}. However, the maximum degree of circular polarization that they could achieve was only 5\%, which is significantly less than the highest value reported in exfoliated samples\cite{mak2,nihit_2defect,cao_vp}.               

Here, we study the effect of systematic \textit{in situ} vacuum annealing and strain relaxation on the luminescence as well as the valley polarization properties of large area strictly 1L-\mos films, grown on sapphire and \sio substrates by a micro-cavity based CVD technique. Investigation using temperature dependent PL, polarization resolved PL and Raman spectroscopy in controlled high vacuum environment, reveals the important roles played by the strain and the $V_S$ defects that are attached with physisorbed air molecules in shaping the optical properties of the CVD grown 1L-\mos. Removal of air molecules from the $V_S$ sites has been shown to enhance the relative intensity of the A-exciton/trion feature with respect to the BL band arising from those defects at low temperatures. It has also been observed that the desorption of these molecules from the $V_S$ defects assists in improving the valley polarization property  of the films. As grown 1L-\mos film on sapphire substrate is transferred using a polystyrene (PS) assisted process on to a \sio substrate for investigating the effect of strain relaxation on the optical properties of 1L-\mos. The transfer is found to result in the relaxation of the biaxial tensile strain in the layer, which in turn helps in almost fully recovering the polarization character of the valleys. It has also been observed that the transfer process enables the removal of physisorbed air molecules from the $V_S$ sites and thus the relative yield of A-exciton/trion transition with respect to the BL band is enhanced. Moreover, the \mos film gets a long lasting PS capping that protects it from the ambient conditions. 

Large area monolayer (1L) \mos films were grown on \textit{c}-sapphire (sample P) and $\sim$300~nm \sio (sample Q) substrates by the microcavity based CVD technique, reported earlier by our group\cite{pkm,tailoring}. Prior to the growth, substrates were  ultrasonically cleaned in trichloroethylene, acetone and methanol successively for 10 minutes each. Sapphire substrate was further cleaned by dipping in a H$_2$O:HF (10:1) solution for 40~s. Another sample, \enquote*{P*} was prepared by transferring 1L-\mos from the sapphire substrate, where the film was grown originally, to a \sio wafer. We used a surface-energy assisted transfer method\cite{transfer,pkm_transfer}, in which polystyrene was first spin coated on the as-grown sample and later baked in two steps; at 80-90~$\degree$C for 35 minutes followed by 120~$\degree$C for 10 minutes. A water droplet was allowed to get into a gap between the sapphire and the polystyrene coating that results in a gradual removal of the whole polystyrene plus \mos assembly from the sapphire substrate. The assembly was then picked up and placed on top of a \sio substrate before baking it again following the same two step baking recipe. Finally, the sample is dipped in toluene several times to remove the polystyrene coating. More details about the process can be found in the supplementary figure~S1. Vacuum ($\sim$10$^{-6}$ mbar) annealing of the samples was performed inside a liquid nitrogen cryostat at 400~K for 30 minutes. Photoluminescence (PL) and Raman measurements were \textit{in situ} carried out on these samples in backscattering configuration at different temperatures inside the same cryostat  before and after annealing. A home-built microscope equipped with a 50$\times$ long working distance objective (NA 0.5) was utilized for these measurements. A 532~nm (2.33~eV) diode pumped solid state laser (DPSS) was used as the excitation source. Signals were recorded by a 0.55~m focal length monochromator attached with a Peltier cooled CCD detector. Polarization resolved PL measurements  were also carried out in the same setup, where a linearly polarized 633~nm (1.96~eV) HeNe laser was used as the excitation source. An achromatic quarter waveplate was used to make the excitation left /right circularly polarized ($\sigma^-$/$\sigma^+$). Another quarter waveplate and a Glan-Taylor analyzer were used to selectively choose left ($\sigma^-$) and right ($\sigma^+$) circularly polarized photons in the collection path of the PL signal (see supplementary Fig.~S2). The degree of circular polarization of the collected PL is defined as $P$ = $(I^{+}-I^{-})/(I^{+}+I^{-})$, where $I^{\pm}$ corresponds to the intensity of $\sigma^{\pm}$ polarized lights. Intensities of the laser spots on the samples were less than 100~$\mu$W and 20~$\mu$W, respectively, for the 532 and 633~nm lasers. Atomic force microscopy (AFM)  was carried out at ambient condition on these samples in tapping mode using Bruker, NanoScope-IV system.


Fig.~\ref{fig1}(a) compares room temperature micro-Raman spectra for all samples. In all cases, the two characteristic vibration modes, namely the in-plane E$^1_{2g}$ and out-of-plane A$_{1g}$ features, are clearly evident. Separation ($\Delta\nu$) between the two peaks is found to be $\sim$20.0, 20.2 and 19.5~cm$^{-1}$ for samples P, Q and P*, respectively. These values of $\Delta\nu$ are well within the range for monolayer \mos suggesting that all the samples  investigated here are indeed 1L-\mos \cite{li_raman,pkm,tailoring} . It should be noted that the E$^1_{2g}$ peak for sample Q (\sio grown) and P* (transferred on \sio) appears almost at the same position  [$\sim$383.2~cm$^{-1}$] while, it shows up at $\sim$384.4~cm$^{-1}$ for the sapphire grown sample P. This may suggest the existence of more biaxial strain in sample P as compared to other two\cite{tailoring,amani_dubey1,baek_CVD}. A$_{1g}$ peak is also down shifted  by  $\sim$1.0~cm$^{-1}$ and $\sim$1.7~cm$^{-1}$, respectively, for sample Q and P* as compared to that of sample P. This down-shift is likely to be also associated with strain relaxation in the two samples Q and P*. The reason for some extra amount of shift in sample P* might be the increase of free electron concentration in the conduction band\cite{sood_raman3}. 

\begin{figure}[h!]
	\includegraphics[scale=1.5]{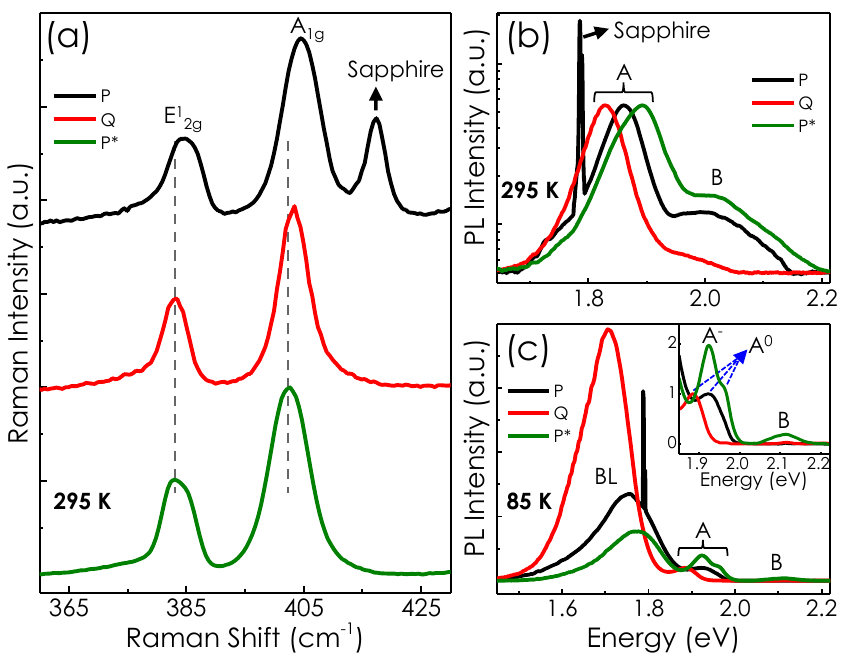}
	\caption{(a) Room temperature Raman spectra normalized at A$_{1g}$ peak for all samples.  PL spectra  recorded at (b) room temperature and (c) 85~K for all samples. Inset of (c) shows the excitonic transitions in expanded scale. PL spectra are normalized  at A exciton (A$^0$) peak. 532~nm laser light is used as the excitation source for these measurements.}
	\label{fig1}
\end{figure} 

Fig.~\ref{fig1}(b) compares the room temperature normalized PL spectra for these samples, which are featured by two peaks. The most dominant peak  appearing at $\sim$ 1.86~eV and a hump at $\sim$ 2.0~eV can be assigned to A and B excitons, respectively. The two excitons result from spin-orbit splitting of the valance band maxima in 1L-\mos\cite{mak2}. Two sharp spikes at around 1.79~eV for sample P are stemming from sapphire substrate. It should be noted that all the spectra in panel (b) can be deconvoluted with three Gaussian functions, of which  two are needed to fit the A exciton feature. The two can be identified as neutral A$^0$ exciton and a red shifted charged A$^-$ trion\cite{mak2,mak_trion,tailoring}. It is evident that the A exciton complex is blue shifted (by $\sim$30~meV) and broadened for the transferred sample P* as compered to the sapphire grown sample P.  This suggests a relaxation of a biaxial tensile strain in the \mos layer after the transfer. It is plausible that such a strain could be developed in 1L-\mos films grown on sapphire substrates due to lattice and/or thermal expansion coefficient mismatch\cite{heteroepitaxy,heteroepitaxy1,dumcenco_triangle}. It is well known that biaxial tensile strain can downshift the A excitonic peak position\cite{amani_dubey1, zhu_vp_strain,conley_raman}. The increased broadening in the transferred sample can be attributed to the increment in trion weightage as a result of increased electron concentration, which is also consistent with the results of Raman measurements shown in Fig.~\ref{fig1}(a). Reason for the enhanced carrier concentration in sample P* will be discussed later. Evidently, both the excitonic peaks  (A and B-excitons) for the \sio grown sample Q  are red shifted by $\sim$30~meV with respect to those for sample P. Note that the position of E$^1_{2g}$ Raman peak for sample Q is almost the same as that of the transferred sample P* and also matches very well with the reported values for the exfoliated samples\cite{li_raman,conley_raman}. This rules out the presence of strain in sample Q. PL peak shift for this sample thus can not be attributed to strain. This point will be discussed later. 

Fig.~\ref{fig1}(c) plots 85~K normalized PL spectra for these samples. In addition to band edge A ($\sim$ 1.95~eV) and B ($\sim$ 2.1~eV) excitons, another broad luminescence (BL) feature appears in these spectra at $\sim$ 1.75~eV, which has been attributed to the excitons bound to $V_S$ defects that are attached with air molecules (O$_2$ and/or N$_2$)\cite{tongay_defect}. Note that the BL feature is earlier been shown to be resulting from excitons bound to S-vacancy related defects in these CVD grown 1L-\mos samples\cite{tailoring,surrente_CVD_vp}. In another report, it has been further demonstrated that the attachment of oxygen molecules with $V_S$ defects passivates its shallow donor nature leading to the reduction of the free carrier concentration in these films\cite{therm}. Note that in case of the transferred sample P*, contributions from  A$^0$ excitons ($\sim$ 1.96~eV) and the trions (A$^-$ at $\sim$ 1.92~eV)\cite{mak2,mak_trion} are quite distinguishable from each other as can be seen in the inset of the figure~\ref{fig1}(c). This suggests that the density of free electrons in the sample is higher than in two other samples as the trion formation probability is expected to rise  with the density of free carriers\cite{amani_dubey1,mouri_Ch,mak2}. This is consistent with the observation of Fig.~\ref{fig1}(a). Another interesting point is that the intensity of BL with respect to that of the A excitonic feature is reduced in the transferred \mos film (sample P*) as compared to the as-grown sample P. This may suggest that the number of attached oxygen molecules (O$_2$) with $V_S$ and hence the passivation of the $V_S$-donors has been reduced in sample P*, which is also in accordance with the  observation of enhancement of free electron concentration in the sample with respect to sample P.  It should be noted that the BL to A-exciton intensity ratio is the maximum for \sio grown sample Q indicating higher density of passivated S-vacancies.  It is plausible that the reduced density of  free electrons in this sample can result in the suppression  of screening of the electron-hole coulomb interaction potential leading to an enhancement in the binding energy of excitons\cite{chow_defect}. This can explain red-shift of all PL peaks in this sample.   
 
\begin{figure}
	\includegraphics[scale=1.5]{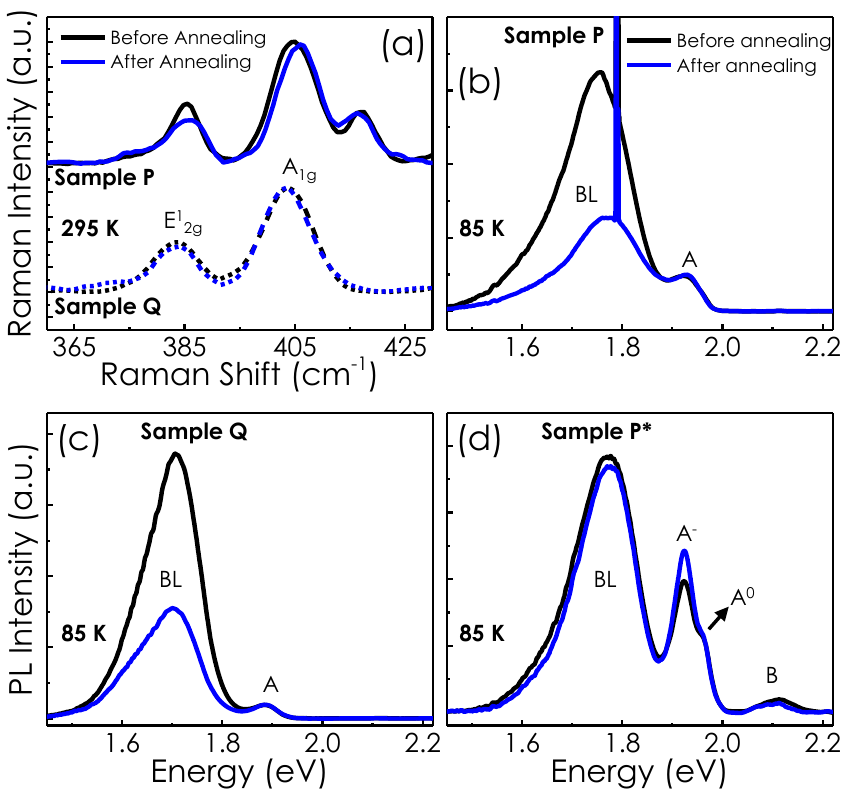}
	\caption{(a) Room temperature normalized (at A$_{1g}$ peak) Raman spectra of sample P (solid lines) and sample Q (dashed lines) before (black) and after (blue) annealing. Normalized (w.r.t.  A$^0$ peak) PL spectra recorded at 85~K, before (black) and after (blue) annealing for sample (b) P, (c) Q and (d) P*. 532~nm laser light is used as the excitation source for these measurements.}
	\label{fig2}
\end{figure}
                  
Fig.~\ref{fig2}(a) compares room temperature Raman spectra for sapphire grown samples P and \sio grown sample Q before and after annealing. It can be seen that for both the samples, Raman peaks before and after annealing overlap with one another which indicates that annealing at 400~K hardly changes the strain state of the materials. Panel (b), (c) and (d) plot normalized (w.r.t. A$^0$ peak) PL spectra at 85~K before and after annealing for samples P, Q and P*, respectively. Note that for the as-grown samples P and Q the intensity of the BL feature substantially reduces after annealing. This can be attributed to the desorption of ambient O$_2$ and/or N$_2$ molecules from $V_S$ cites\cite{tongay_defect}. However, for the transferred sample P*, where BL intensity is already quite reduced, there is not much of change in BL intensity upon annealing. This result may indicate that most of the S-vacancies are already depleted in this sample. It should also be noted that in sample P*, there is a slight tendency of increment of relative intensity of A$^-$ peak with respect to that of A$^0$ upon annealing.  As mentioned earlier that in one of our recent publications, physisorption of oxygen molecules at $V_S$-sites is shown to be responsible for the passivation of the $V_S$ donors leading to the reduction of the free carrier concentration in these films\cite{therm}. Observation of annealing driven increase of trion intensity in sample P* could thus mean that some of the left-over oxygen molecules (O$_2$) from the $V_S$-sites are removed after annealing leading to the enhancement of density of free electrons in the layer. However the number of such removals is not enough for giving rise to any significant change in BL intensity.


\begin{figure}[h!]
	\includegraphics[scale=1.5]{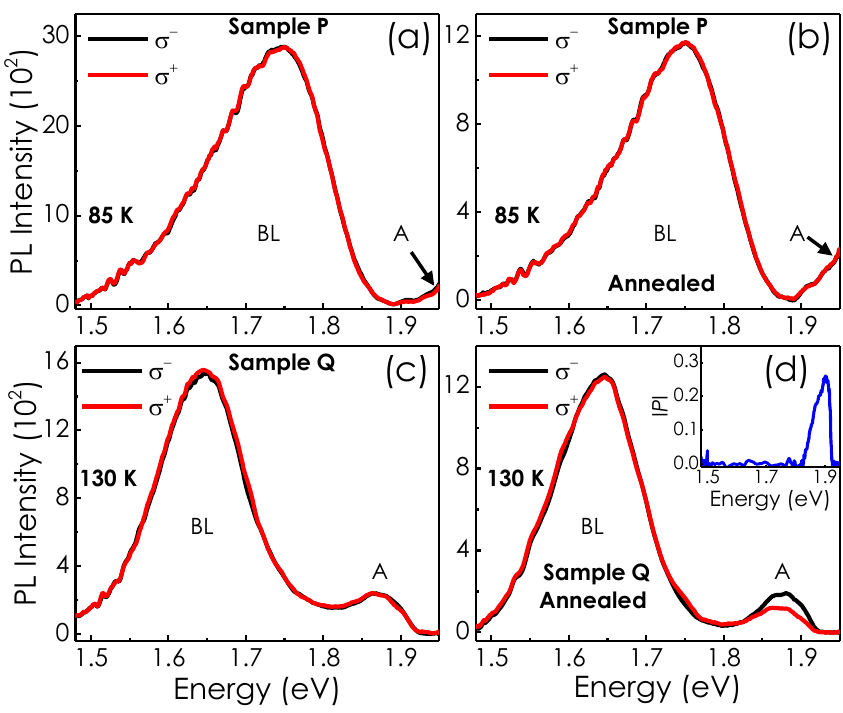}
	\caption{Circular polarization resolved PL spectra for sample P recorded at 85~K (a) before and (b) after annealing. Same PL spectra for sample Q recorded at 130~K (c) before and (d) after annealing. Inset of (d) shows degree of circular polarization ($|P|$) as a function of photon energy. Black (red) spectra in all panels are obtained for $\sigma^-$ ($\sigma^+$) polarized luminescence when $\sigma^-$ polarized  light of 633~nm laser is used for excitation.}
	\label{fig3}
\end{figure}

Fig.~\ref{fig3}(a) and (b) show circular polarization resolved PL spectra recorded at 85~K temperature for sapphire grown sample P before and after annealing, respectively, where left circularly polarized ($\sigma^-$) 633~nm laser (1.96 eV) is used for excitation. Evidently, $\sigma^-$ and $\sigma^+$ polarized PL spectra are overlapping with one another before and after annealing. Fig.~\ref{fig3}(c) and (d) show polarization resolved PL spectra obtained with $\sigma^-$ polarized 633~nm laser excitation at 130~K temperature for sample Q before and after annealing, respectively. While there is no change between the spectra recorded for different polarizations for the sample before annealing, a clear difference emerges at the A-exciton/trion feature after annealing. Inset of Fig.~\ref{fig3}(d) plots absolute value of degree of circular polarisation ($|P|$) as a function of photon energy. It is noteworthy that as high as 25$\%$ polarization can be achieved at the peak position of A$^0$. It should be noted that we have performed the same experiments at 85~K on sample Q before and after the annealing. However, at that measurement temperature, A-exciton/trion feature could not be seen and the BL feature that dominates the spectrum is found to be unpolarized both before and after the annealing(see supplementary Fig.~S3). At this low temperature, exciton capture rate at the $V_S$ cites must be much higher than the recombination rate of the free excitons as the energy and power of the 633~nm laser is relatively low\cite{tongay_defect}. BL feature, which are likely to be originating from the excitons trapped in $V_S$-plus-ambient-molecule complexes, has not been reported to show polarization in 1L-\mos\cite{mak2,nihit_2defect}. Excitons trapped in these defects are expected to have a spread in their crystal momentum, which can facilitate spin flip scattering between K and K$^\prime$ valleys through Bir-Aranov-Pikus electron-hole exchange mechanism resulting in a mix-up of excitons with opposite spins\cite{mak2, yu_vdepol_eh}.  

 \begin{figure}
 	\includegraphics[scale=1.5]{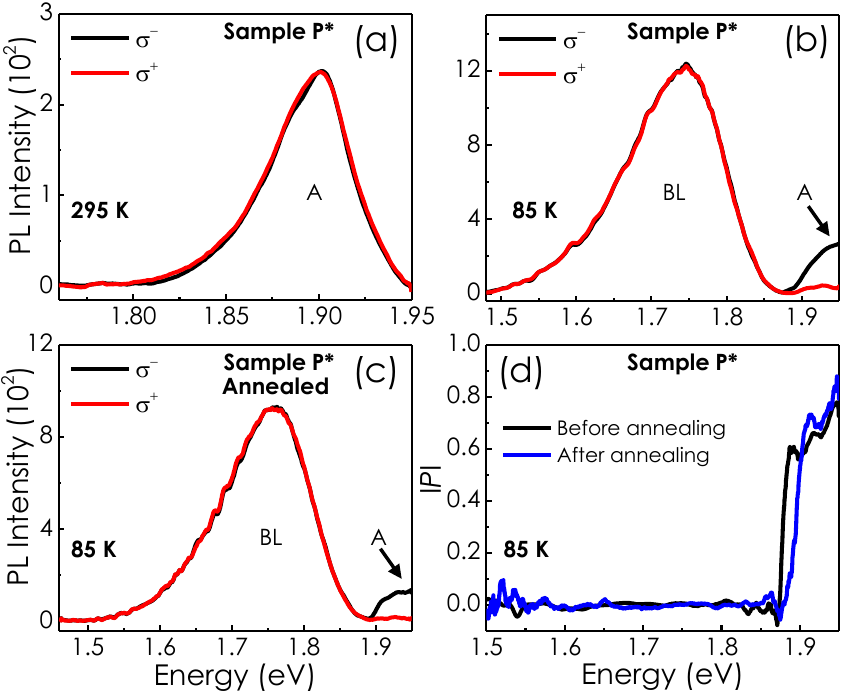}
 	\caption{Circular polarization resolved PL spectra for the transferred sample P* recorded at (a) 295~K, (b) 85~K before annealing and (c) 85~K after annealing.  $\sigma^-$ polarised 633~nm laser light is used as the excitation source. (d) Degree of circular polarization ($|P|$) as a function of photon energy before (black) and after (blue) annealing.}
 	\label{fig4}
 \end{figure}

Fig.~\ref{fig4}(a) and (b) compare $\sigma^-$ and $\sigma^+$ polarized  PL spectra recorded at 295~K and 85~K, respectively, for the transferred sample P*. While at 295~K, $\sigma^-$ and $\sigma^+$ profiles overlap with one another, a gap clearly opens up at A-exciton/trion feature at 85~K. Note that gap opening could not be found at the BL feature. Panel (c) shows the comparison between $\sigma^-$ and $\sigma^+$ polarized PL spectra for the sample P* after annealing. It is noticeable that the gap opening at the A-exciton/trion feature is still intact even after the annealing. Fig.~\ref{fig4}(d) plots the degree of circular polarization ($|P|$) as a function of photon energy for the sample before and after the annealing. It is remarkable to see that for both the cases, $|P|$ reaches up to $\sim$ 0.8 at the A$^0$ exciton position. 

Note that in these experiments, layers are excited with a circularly polarized laser light whose energy is in resonance with A exciton complex at 85~K temperature. One should thus expect $\sim$ 100$\%$ PL-polarization at A-exciton/trion feature in case of an ideal 1L-\mos as K and K$^\prime$ valleys can only accommodate excitons with pseudospin +1 and -1, respectively\cite{asakura_coherence}. $|P| < 1$ at A-exciton/trion feature should thus mean a deviation from ideality. It is conceivable that the presence of strain and defects in the layer can influence the value of $|P|$  by introducing spin-flip channels through band structural modification \cite{zhu_vp_strain,raman2,zeng_vp1}. In fact, it has been experimentally shown that the enhancement of the uniaxial tensile strain can lead to the reduction of $P$ in exfoliated \mos monolayers\cite{zhu_vp_strain}. Theory predicts that the increase of uniaxial\cite{zhu_vp_strain,conley_raman} as well as biaxial\cite{scalise_biaxial} tensile strain, leads to the up shift of the  energy of the spin degenerate local valance band maximum (VBM) at the $\Gamma$ point ($\Gamma_v$) in 1L-\mos.  Beyond a threshold value of strain, the $\Gamma_v$-maximum  moves above the VBM at K(K$^\prime$) and the band gap becomes indirect. This tensile strain driven upward movement of VBM at the $\Gamma$-point  increases the probability of spin flipping via spin-degenerate $\Gamma$ valley\cite{zhu_vp_strain,mai_manybody}. This is one of the reasons for valley depolarization of 1L-\mos. Here, the sapphire grown sample P has been found to posses the highest amount of tensile strain. Observation of zero PL-polarization in sample P both before and after annealing can thus  primarily be attributed to the strain. The findings of Fig.~\ref{fig1} clearly suggest the release of tensile biaxial strain in transferred sample P*. Relaxation of strain might thus be a reason for the recovery of the polarization in the transferred film. Note that tensile biaxial strain in \sio grown sample Q is much lower than that of the sapphire grown sample P [see Fig.~\ref{fig1}]. So, the observation of zero PL-polarization in sample Q before annealing cannot be attributed to the modification of band structure due to strain. Note that the density of S-vacancy defects and hence of $V_S$-plus-air-molecule complexes is found to be more in sample Q than in other samples [see Fig.~\ref{fig1}(c)] before annealing.  Upon annealing, when air molecules are removed from $V_S$ sites, the intensity of BL feature decreases and at the same time sample Q starts showing a finite $|P|$ at the A-exciton/trion feature. This suggests that air molecules adsorbed at the  $V_S$ sites might be responsible for nullifying $|P|$ in this sample. Among different elements in air,   O$_2$  possesses a magnetic moment of 2.0~$\mu_B$ per molecule\cite{pauling_o2_magent}. One can thus expect the existence of a finite magnetic moment also at the $V_S$-plus-O$_2$-molecule defect sites. For an example, density functional theory (DFT) calculations predict a magnetic moment of $\sim$ 1.8~$\mu_B$ for the O$_2$ molecule when it is adsorbed on the surface of $\alpha$-alumina (0001)\cite{wang_o2_magnet}. As mentioned earlier, physisoption of oxygen molecules at the $V_S$ sites leading to the passivation of of these defects as donors has been demonstrated for these samples\cite{therm}.  Therefore, it is reasonable to believe that  O$_2$ molecules attached to the$V_S$-sites can act as  spin-flip centres, which promote  the inter valley (K-K$^\prime$) spin-flip scattering of the excitons/trions in \mos.  Note that  the suppression of BL-intensity in the transferred sample (sample P*) as compared to the 1L-\mos film grown on sapphire substrate (sample P) as shown in Fig.~\ref{fig2} suggest a reduction of the density of the O$_2$ molecules adsorbed at the $V_S$-sites. Like in case of \sio grown sample Q (see Fig.~\ref{fig4}), removal of these molecules from the S-vacancy sites might also contribute to the recovery of $|P|$ in transferred sample P*.

\begin{figure}
	\includegraphics[scale=1.5]{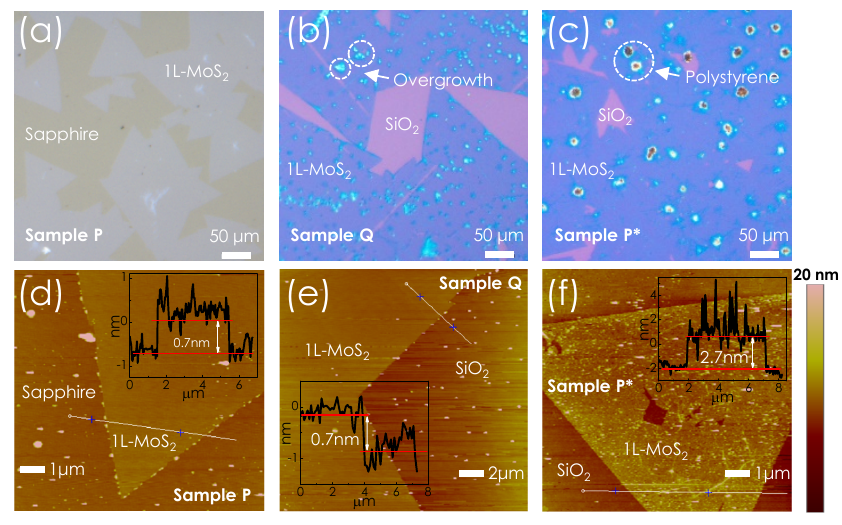}
	\caption{Optical microscopic images of samples (a) P, (b) Q and (c) P*. AFM topographic images of samples (d) P, (e) Q and (f) P*. Insets of panels (d), (e), and (f) show the height distribution along the lines drawn across the boundaries of samples P, Q and P*, respectively.}
	\label{fig5}
\end{figure}

It should be noted that after the suppression of the BL-intensity through annealing, if the as-grown samples P and Q are exposed to the ambient condition, the relative intensity of BL with respective to that of A exciton/trion feature has been found to return to its original value after several hours of exposure. This reversibility suggests that the air molecule are getting physically\cite{therm,tongay_defect} adsorbed and desorbed from the $V_S$-sites in these films. However, similar experiment on the transferred sample P* does not show  much variation of BL-intensity even after keeping the sample in the ambient for several months as shown in supplementary figure S4. Note that the transfer process involves coating \mos with polystyrene followed by baking at temperature comparable to that is used for vacuum annealing. It could be possible that the transfer process itself is facilitating the removal of the adsorbed molecules from the $V_S$-sites and then provides an encapsulation of the layer that protects $V_S$-centres from air. Fig.~\ref{fig5}(a-c) compare the optical microscopic images for samples P, Q and P*. \mos films covering large areas can be found for all the samples. Layer grown on sapphire substrate looks quite clean, whereas some small high contrast triangular overgrown structures can be seen on \mos layer grown on \sio substrate. However, a much higher density of circular structures can be seen on transferred sample P* [panel (c)]. These structures can be attributed  to some bigger residues of polystyrene\cite{transfer}, which could not be dissolved completely in toluene at the final stage of the transfer process. Fig~\ref{fig5}(d-f) show AFM micrographs for the three samples. Clear contrast between the \mos flakes and the substrate is evident in these images. Average height of the flakes can be obtained from the line scan profiles recorded across the boundary between \mos and the substrate at several places. Inset of each of the panels shows one such profile recorded along the white line shown in the image. For the samples P and Q , the average step height of the flakes has been found to be $\sim$0.7~nm. This ensures monolayer nature of the \mos flakes in both the samples. Average roughness of these flakes is found to be $\sim$0.12~nm. In case of the transferred sample P*, the average step height and the roughness of the flakes come out to be $\sim$2.7~nm and $\sim$0.89~nm, respectively. Both these values are clearly larger than those are obtained for the as grown samples P and Q, which clearly suggests that a thin capping layer still exists on \mos in the transferred sample P* even after cleaning the sample several times in toluene. All these findings strongly support our conjecture that the transfer process not only help in removing the physisorbed air molecules from \mos, but also protect the layer from the ambient by introducing a coating on the surface.


In conclusion, vacuum annealing helps in removing air molecules from sulfur vacancy sites in the CVD grown 1L-\mos that leads to the suppression of the broad luminescence (BL) feature that results from sulfur vacancy ($V_S$)-air-molecule defect complexes at low measurement temperatures. In case of sapphire grown 1L-\mos sample, no difference between $\sigma^-$ and $\sigma^+$ polarized PL spectra could be found in the whole temperature range of investigation (85 to 300~K) when circularly polarized 633~nm laser light is used for excitation. Degree of circular polarization $|P|$, which has been defined as the ratio between the change in $\sigma^-$ and $\sigma^+$ polarized PL intensity and the overall PL intensity at a given photon energy, is found to be zero throughout the spectra even after the sample is vacuum annealed. On the other hand, in case of sample P*, where 1L-\mos layers grown on the sapphire substrate are transferred on a \sio substrate, $|P|$ at A-exciton/trion position is found to be as high as 0.8. This shows that the main cause for showing zero polarization at A-exciton/trion feature in case of the sapphire grown sample even after the annealing is likely to be the biaxial tensile strain, which is shown to be present in the as-grown layer. In case of the as-grown sample on \sio substrate, $|P|$ is found to be zero over the entire photon energy range as well as the temperature range of investigation. However $|P|$ at the A-exciton/trion position becomes as high as 0.25 after the vacuum annealing  of the sample. This suggests that the recovery of polarization at the A-exciton/trion position upon annealing must be associated with the removal of air molecules from the $V_S$ sites. The study thus shows that the strain as well as the physisorption of air molecules with $V_S$ sites play key roles in governing the valley polarization property of CVD grown 1L-\mos. It has also been found that the polystyrene assisted process adopted here for transferring CVD grown \mos monolayers from sapphire substrates to \sio substrates not only facilitates the removal of physisorbed air molecules from the $V_S$ sites but also shields the \mos film from the environment by putting in place a long lasting capping layer. The transfer technique thus enhances the relative yield of A-exciton/trion transition by suppressing the BL transition. The method also  significantly improves the valley polarization property of the film. Both the improvements are highly desirable for the development of  valleytronic devices in the future. 
\\ 
\vspace*{-\baselineskip}


\noindent\textbf{Acknowledgment:}
We would like to acknowledge the financial support by Department of Science and Technology (DST) of Government of India (grant number: CRG/2018/001343). We are also thankful for various experimental opportunities provided by Industrial Research and Consultancy Centre (IRCC) and Sophisticated Analytical Instrument Facility (SAIF) of IIT Bombay. We sincerely thank Prof. Arindam Chowdhury for valuable discussions and help in doing certain experiments.  

\bibliographystyle{unsrt}
\bibliography{BIBManuscript}

%





\noindent
{\huge Supplementary Information}
\\

\noindent
\textbf{{\Large{Effect of strain and sulfur vacancies on the luminescence and valley polarization properties of CVD grown monolayer MoS$_2$ films }}}
\\
{\large Poulab Chakrabarti*, Faiha Mujeeb, Subhabrata Dhar}\\
Department of Physics, Indian Institute of Technology Bombay, Mumbai-400076, India

\date{\today}
\maketitle

\clearpage

\begin{figure}
	\includegraphics[scale=0.7]{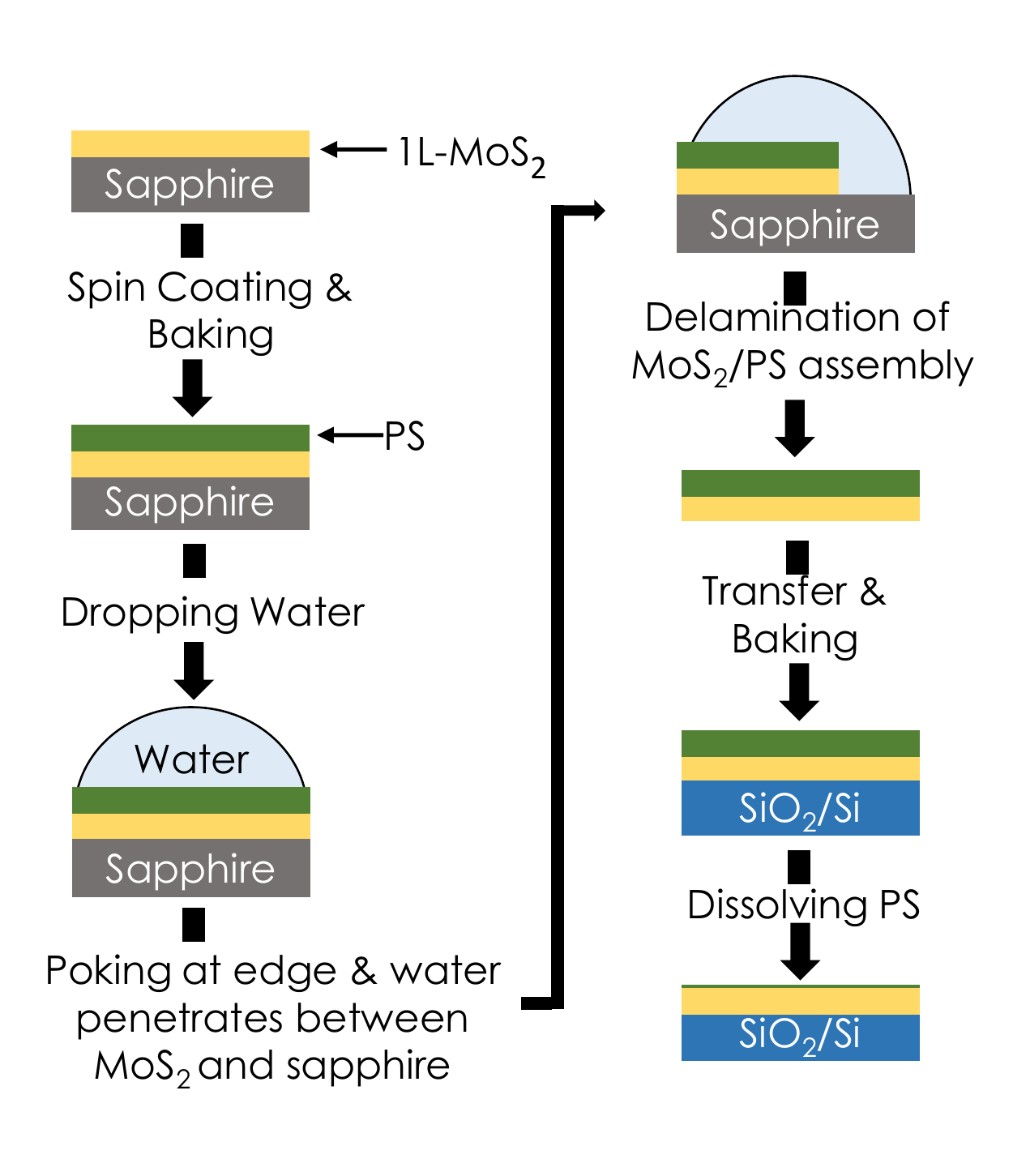}
	\centering
	\caption{The surface-energy assisted transfer process\cite{transfer,pkm_transfer} of 1L-\mos is schematically shown. Here, Polystyrene (PS) was used as the carrier polymer and its solution was prepared by dissolving 9~g of PS with a molecular weight of 280,000~g/mol in 100 ml of toluene. The solution was then spin-coated (500 rpm for 30 s and 3500 rpm for 60 s) on 1L-\mos grown on sapphire and baked at 80-90~$\degree$C for 35 minutes and 120~$\degree$C for 10 minutes, successively. Then a water droplet was dropped onto the PS layer. Water has a natural tendency to penetrate between the \mos film and the substrate due to the differences in their surface energies, but the penetration actually cannot start by itself. Gentle poking at the edge of the \mos/PS assembly was required to initiate the water penetration underneath \mos layer. As a result, the \mos/PS assembly got lifted and it was transferred onto SiO$_2$/Si substrate. After that, it had to be baked for 80-90~$\degree$C for 35 minutes and 120~$\degree$C for 10 minutes. Finally, the PS layer was removed by several times rinsing it in toluene.}
\end{figure}
\clearpage
\begin{figure}[h!]
	\includegraphics[scale=0.8]{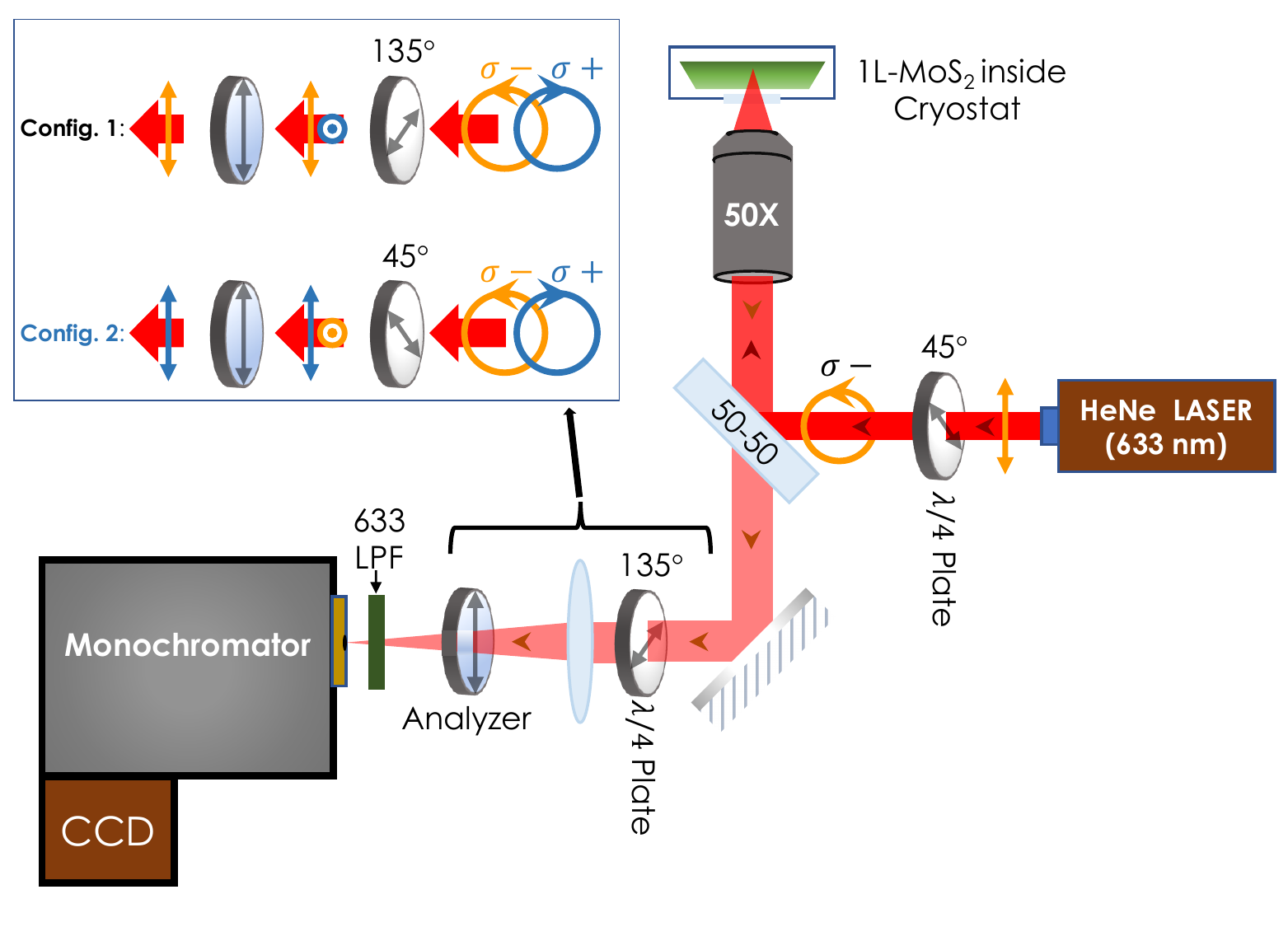}
	\centering
	\caption{Schematic depiction of the circular polarization selective micro photoluminescence (PL) setup. The sample was kept inside a pour-fill type liquid nitrogen cryostat under high vacuum at $\sim$10$^{-6}$ mbar pressure where the sample temperature can be varied from 85~K to 400~K. A linearly polarized Helium Neon (HeNe, 633 nm) laser is made left circularly polarized ($\sigma-$) by a 45$\degree$ oriented achromatic quarter waveplate and used as the excitation source. The beam was then focused on the sample through a quartz glass window, using a 50$\times$ long working distance objective of 0.5 numerical aperture (NA). The collection of backscattered PL signal was also done by the same objective lens and it was further passed through another quarter waveplate and a Glan-Taylor analyzer placed in front of the entrance slit of the monochromator. As shown in the inset of this figure, the quarter waveplate is rotated by 90$\degree$ to selectively choose $\sigma-$ (config. 1) and $\sigma+$ (config. 2) photons which is detected by a Peltier cooled CCD attached to the 0.55~m focal length monochromator. A 633 nm long pass filter (LPF) is placed in front of the monochromator entrance to block the elastically scattered laser light. The same setup is used for Raman spectroscopy, where the HeNe laser and the 633 nm LPF are replaced by a 532 nm diode pumped solid state (DPSS) laser and a Raman compatible ultra sharp 532 nm long pass filter, respectively.}
	\label{Fig:2}
\end{figure}

\begin{figure}
	\includegraphics[scale=1.5]{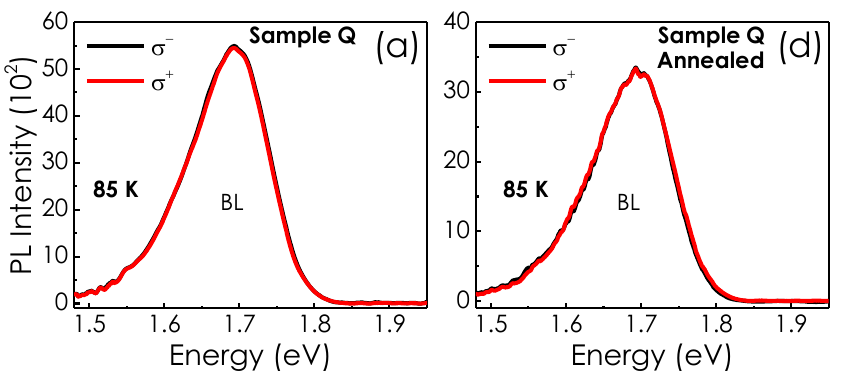}
	\centering
	\caption{Circular polarization resolved PL spectra for as-grown \mos on SiO$_2$/Si (sample Q) recorded at 85~K (a) before and (b) after annealing, where left circularly polarized ($\sigma^-$) 633~nm laser is used for excitation. In both the cases, A-exciton/trion feature is absent and the spectra are dominated only by the $V_S$ related BL peak. As discussed in the main text, here also, BL feature does not show any polarization contrast before and after annealing. The absence of A-exciton/trion feature can be attributed to the low excitation power and energy of the 633~nm laser\cite{tongay_defect}, at which  the radiative recombination rate of the free excitons is overwhelmed by that of the excitons captured at the V$_S$-sites at low temperatures.}
\end{figure}

\begin{figure}
	\includegraphics[scale=1.4]{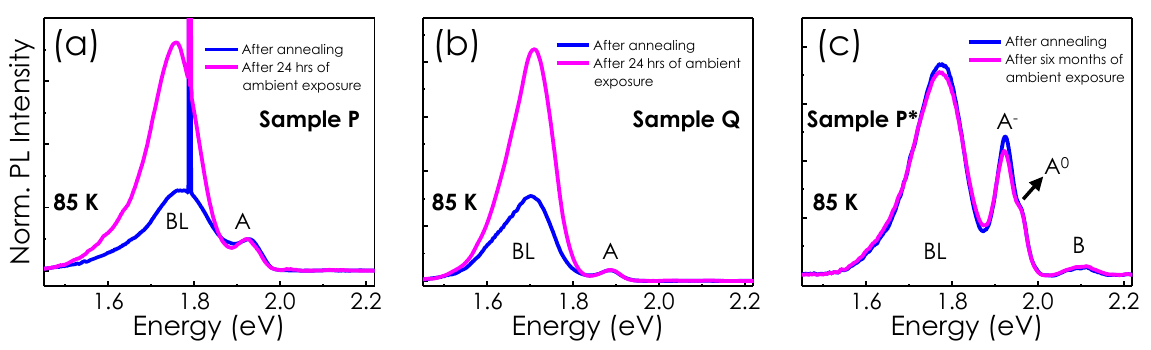}
	\centering
	\caption{Blue curves in panel (a), (b) and (c) show normalized (at A$^0$ peak) PL spectra at 85~K for samples P, Q and P*, respectively, recorded after annealing at 400~K temperature for 30 minutes in high vacuum ($\sim$10$^{-6}$ mbar) condition. After that, the vacuum was broken at room temperature and all the samples were exposed to ambient condition for 24 hours. Again as grown sample P and Q were taken down to 85~K temperature and the normalized (at A$^0$) PL spectra for these samples are plotted in magenta color in panel (a) and (b) respectively. It can be seen that the intensity of BL feature with respect to that of A$^0$ peak has increased and returned to its pre-annealed value (see figure 2 of main text) for sample P and Q. Interestingly for transferred sample P*, the intensity of BL feature relative to A$^0$ peak remains constant after 24 hours exposure of the sample in ambience. Finally, the transferred sample P* was kept in ambient condition for 6 months. The magenta curve in panel (c) is showing the low temperature (85~K) normalized (at A$^0$) PL spectrum of sample P* after 6 month of ambient exposure and it can be seen that it has hardly changed. So, the polystyrene coating gives a  long lasting protection to the sample from air.}
\end{figure}


\end{document}